\documentclass[onecollarge,natbib]{svjour2}
\bibpunct{[}{]}{;}{n}{}{,} 
\smartqed  
\usepackage{graphicx,amsmath}
\journalname{Few-Body Systems (LC2016)}
\def\be{\begin{eqnarray} &&}

\def\nonu{\nonumber \\ &&}
\def\ee{\end{eqnarray}}
 \usepackage{wrapfig}
 \usepackage[normalem]{ulem}
 \usepackage{graphicx,color}
 \usepackage{enumerate}

   \font\bi=cmmib10 scaled 1200         
     \font\tenbifull=cmmib10 scaled 1200 
     \font\tenbimed=cmmib9
     \font\tenbismall=cmmib7
\def\bi{\begin{itemize} }
\def\ei{\end{itemize} }
\def\be{\begin{eqnarray} &&}
\newcommand{\psla}{ \slash \! \! \!}

\def\nonu{\nonumber \\ &&}
\def\ee{\end{eqnarray}}

\makeindex

\def\beq{\begin{equation}}
\def\eeq{\end{equation}}
\usepackage{amsbsy,amsmath}

\newcommand{\bq}{\begin{eqnarray}}
\newcommand{\eq}{\end{eqnarray}}

\begin{document}

\title{ Two-fermion Bethe-Salpeter Equation  in
Minkowski Space: the Nakanishi Way}




\author{Giovanni Salm\`e   \and Wayne de Paula \and Tobias Frederico \and Michele Viviani
 }


\institute{
          Giovanni Salm\`e \at
             INFN, Sezione di Roma, Rome, Italy.
	     \email{salmeg@roma1.infn.it}  \and
         Wayne de Paula \at
           Dep. de F\'\i sica, ITA, S\~ao Jos\'e dos
Campos, S\~ao Paulo, Brazil. \email{wayne@ita.br}\and  Tobias Frederico \at
           Dep. de F\'\i sica, ITA, S\~ao Jos\'e dos
Campos, S\~ao Paulo, Brazil. \email{tobias@ita.br}
                   \and
 Michele Viviani INFN Sezione di Pisa, Pisa, Italy. 
\email{michele.viviani@pi.infn.it}}
\date{}
\maketitle
\begin{abstract}
 The possibility of solving the Bethe-Salpeter Equation in Minkowski space, even for fermionic systems, is becoming actual,
 through the applications of well-known tools: i) the Nakanishi
 integral representation of the Bethe-Salpeter amplitude  and ii) the light-front projection onto the null-plane. 
 The theoretical
 background and some preliminary calculations are illustrated, in order to show the potentiality and the wide range of application
 of the method. 
\end{abstract}

\section{Introduction}
\label{intro}
To achieve a fully covariant description of  two-fermion  systems, directly in
Minkowski space, represents a challenging issue, that nowadays can be considered
quite  conceivable at least for analytic interaction kernels. As a matter of
fact, the description can be carried out within  the non perturbative framework given by 
 the Bethe-Salpeter equation(BSE) \cite{BS51} (see also Ref. \cite{Nakarev} for a wide introduction)
 and exploiting new approaches based on i) the so-called Nakanishi integral representation (NIR)
 \cite{nak63,nak71} of the  BS amplitude (BSA) and ii) the light-front (LF) machinery. For instance, 
 to work directly in the LF environment
  has a clear advantage in hadron physics,  since one is immediately ready to calculate the relevant LF
 momentum distributions to be adopted in the investigation of several processes.
However, it is worth reminding that 
to determine from the BSA, directly in Minkowski space, the LF 
momentum distributions (indeed, to be used
in many areas, besides hadron physics) is the Holy Grail for  both  fundamental approaches, like the lattice
calculations (though in Euclidean space) and  phenomenological studies. 
As it is  well-known, the fermionic nature of the interacting constituents produces
difficulties that are far from trivial, but  in our
approach, where
 the NIR plays a pivotal role in the elaboration of the strategy
  for getting actual solutions of BSE, can be put under control.

\section{The BSE in a nutshell}
\begin{figure}
\begin{center}
\includegraphics[width=12.0cm]{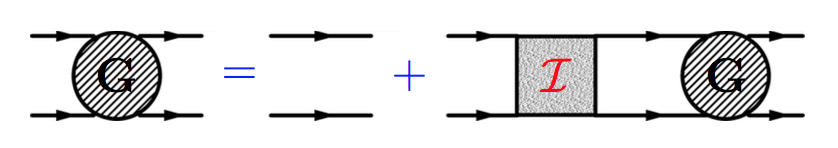}
\caption{ Pictorial representation of the integral equation fulfilled by the
4-point Green's function.}
\label{fig1}
\end{center}
\end{figure}

\begin{figure}
\begin{center}
\includegraphics[width=10.cm]{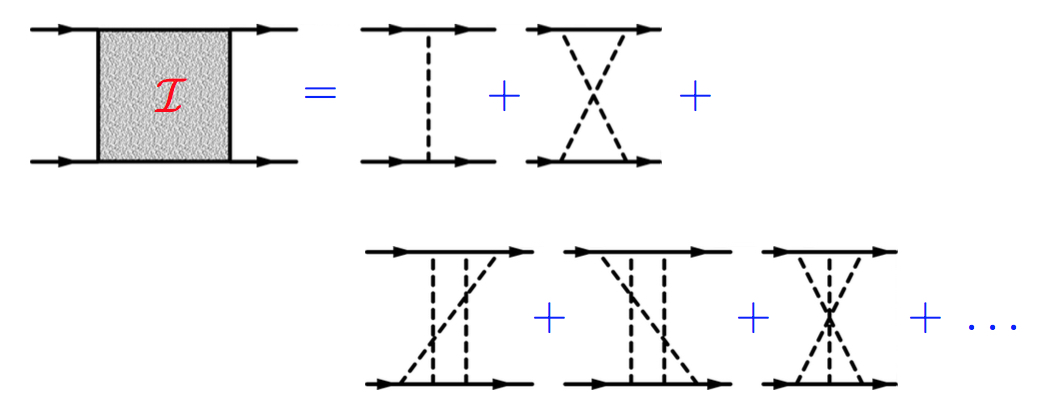}
\caption{First contributions to the interaction kernel in the integral equation
for the 4-leg Green's function (cf Eq. \eqref{g4ie}).}
\label{fig2}
\end{center}
\end{figure}

We briefly recall the main path to BSE (for an extended review see, e.g., 
Ref. \cite{Nakarev}), simplifying to the case of two scalars. One has to start with 
 the 4-point Green's Function, 
given by 
\be G(x_{1},x_{2};y_{1},y_{2})=<0\,|\,
T\{\phi_{1}(x_{1})\phi_{2}(x_{2})\phi_{1}^{+}(y_{1})
\phi_{2}^{+}(y_{2})\}\,|\,0>~~~.\ee 
It fulfills an integral equation  (see  Fig. \ref{fig1}) that symbolically reads as follows
\be
G=G_0~+~G_0~ {\cal I}~G
\label{g4ie}\ee
where ${ \cal I}$ is the interaction  kernel 
  given by the 
infinite sum of irreducible Feynman graphs. In Fig. \ref{fig2} some examples are
given for the simple case of a $\phi^3$ theory (see Ref. \cite{Baym} for the caveats about such a model).
 It should be pointed out that the iteration of the integral equation gives all 
 the expected contributions.
If one  
 inserts a complete Fock basis   in $G(x_{1},x_{2};y_{1},y_{2})$, 
then  the bound state contribution
(assuming only one non degenerate bound state for the sake of simplicity)
appears as a pole in the Fourier space, i.e.
\be G_B(k,q;p_B,\beta)\simeq {i\over (2\pi)^{-4}}~
\frac{{ {\phi(k;p_{B},\beta)~\bar{\phi}(k;p_{B},\beta)}}}
{2\omega_{B}(p_{0}-\omega_{B}+i\epsilon)}
\ee
where $\beta$  is the set of possible quantum numbers,
  $\omega_B=\sqrt{M^2_B+|{\bf p}|^2}$,
  $\phi(k;p_{B},\beta)$ is 
   BSA for a bound state,
 in momentum  space. In configuration space, 
  the BSA reads 
$
\langle0|T\{\phi_{1}(x_{1})\phi_{2}(x_{2})\}
|p_B\,\beta\rangle$.

 Close to the bound-state pole, i.e.  $p_0\to \omega_B$  the 4-point Green's function can be approximated by
$G\simeq G_B+~regular~terms$ and one gets 
the  homogeneous  BSE, that holds for a bound state. In conclusion,
the integral equation determining  the BSA for a bound system,
  without self-energy and vertex corrections to be simply, is given by 
\be   \phi(k;p_{B},\beta)= G_0(k;p_B,\beta) ~\int
d^{4}q'~{  {\cal I}(k,q';p_B)}~\phi(q';p_{B},\beta) \ee
where  for a two-scalar system the two-body free-propagator is
\be
G_0= \frac{i}{(\frac{p_B}{2}+k)^2-m^2+i\epsilon}~~\frac{i}{({\frac{p_B}{2}}-k)^2-m^2+i\epsilon}
\ee
Notably, ${\cal I}(k,q';p_B)$, the irreducible kernel in BSE, 
is the same one meets 
in the integral equation for the 4-point Green's function (cf Eq. \eqref{g4ie}).

\section{ Nakanishi  integral  representation of  $N$-leg transition
amplitudes}
 
In the sixties, Nakanishi \cite{nak63,nak71}
proposed an integral representation for   
$N$-leg transition amplitudes, based on the parametric formula of  the Feynman
diagrams.
\begin{figure}
\begin{center}
\includegraphics[width=5.cm]{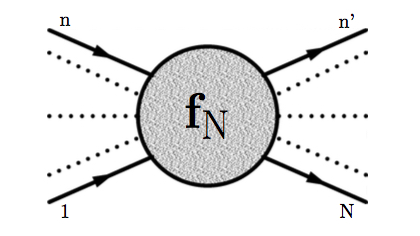}
\caption{Pictorial representation of a  $N$-leg transition amplitude. Notice that the legs
can correspond to on-mass-shell particles or to off-mass-shell ones, depending on the actual process under scrutiny.}
\label{fig3}
\end{center}
\end{figure}
In a scalar theory, for $N$ external legs, a generic contribution to the transition
amplitude is given by (see Fig. \ref{fig3})
 \be
 f_{{  \cal G}_{nk}}(p_1,p_2,...,p_N)\propto ~\prod_{r=1}^k\int d^4q_r~
\frac{1}{(\ell_1^2-m_1^2)(\ell_2^2-m_2^2)~\dots ~(\ell_n^2-m_n^2)}\ee
where  $n$ propagators and $k$ loops are present.
($k$ is the number  
of integration variables). It is important to notice that the external momenta are $N$ and does not change from one diagram to another, all
belonging to the infinite set contributing to the given process one is  investigating.

 Nakanishi proposed  a compact
and elegant expression for the full $N$-leg amplitude
$$f^N(s)=\sum_{{  \cal G}_{nk}} ~ f_{  {\cal G}_{nk}}(s)$$
where $s\equiv\{s_h\}$ is the set of all the independent scalar products one can construct
from the $N$ external momenta.
The main ingredient for obtaining  the Nakanishi integral  representation, once
the Feynman parametric representation for the amplitudes is adopted,
is the following identity
\be
1\doteq \prod_h \int_0^1dz_h
\delta\left(z_h-\frac{\eta_h(\vec \alpha)}{\beta}\right )
\int_0^\infty d\gamma~\delta\left(\gamma-\sum_l\frac{\alpha_lm_l^2}{\beta}
\right)\ee
where $\{\eta_h\}$  are suitable combinations of the Feynman parameters $\vec \alpha\equiv \{\alpha_i\}$, and 
$\beta=\sum_h \eta_h(\vec \alpha)$ \cite{nak63,nak71}. By integrating by parts $n-2k-1$ times
one gets
\be
 f_{{\cal G}_{nk}}(s)\propto~\prod_h\int_0^1dz_h\int_0^\infty 
d\gamma 
\frac{\delta(1-\sum_h z_h)~\tilde{\phi}_{\cal G}(z,\gamma)}
{{ {(\gamma-\sum_h z_hs_h)}}}
\label{nir0}\ee
where  $\tilde{\phi}_{{\cal G}_{nk}}(z,\gamma)$  is a proper function (indeed a
distribution).  
 By adopting the Nakanishi change of variables one is able
 to move  the dependence upon the details of the diagram,  i.e.
 $(n,k)$,  from
the denominator to  the numerator. More notably,  after performing the change of
variables any   diagram ${\cal G}_{nk}$ leads to a contribution $ f_{{\cal G}_{nk}}(s)$ with 
just the  same  formal 
expression for  the denominator, as given in Eq. \eqref{nir0}.  Eventually, this allows one to formally carry on
 the sum over the
infinite set of Feynman diagrams contributing to a given $N$-leg transition
amplitude $f^N(s)$.
 
As a matter of fact, the full  $N$-leg transition amplitude can be
 formally written as
\be f^N(s)~=\sum_{{\cal G}_{nk}} ~ f_{{\cal G}_{nk}}(s)\propto~\prod_h\int_0^1dz_h
\int_0^\infty 
d\gamma 
\frac{\delta(1-\sum_h z_h)~\phi_N(z,\gamma)}{(\gamma-\sum_h z_hs_h)} \ee
where
$ {\phi}_N(z,\gamma)=\sum_{{\cal G}_{nk}}  \tilde{\phi}_{{\cal G}_{nk}}(z,\gamma) $, is 
called the Nakanishi weight function for the $N$-leg transition amplitude.
 
\begin{figure}
\begin{center}
\includegraphics[width=5.cm]{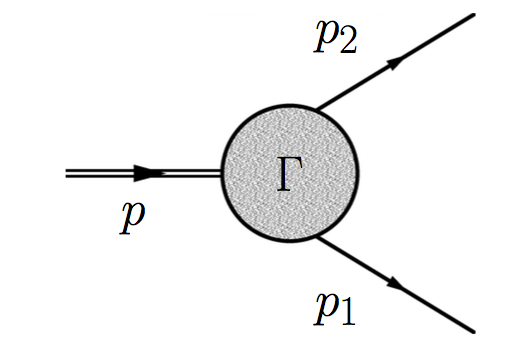}
\caption{The pictorial representation of the BSA, with  $p$ 
 an on-mass-shell 4-momentum.}
 \label{fig4}
\end{center}
\end{figure}

Such an elegant expression can be exploited
for obtaining    
the $3$-leg
transition amplitude or vertex function,  that will be the main quantity to be
tentatively adopted within the BS framework for describing a two-body interacting
system, as discussed in what follows.
The vertex function   reads
\be f_3(s)=\int_0^1 dz 
\int_{0}^\infty d\gamma \frac{\phi_3(z,\gamma)}
{\gamma-{p^2\over 4}- k^2-z k\cdot p-i\epsilon}
\label{vertex1}\ee
with $p=p_1+p_2$ and $k=(p_1-p_2)/2$. Notice that only three
independent scalar products can be constructed from the 4-momenta at disposal. In  Fig. \ref{fig4}, it is pictorially 
shown the BSA, that can be obtained
from $f_3(s) $ after  multiplying   {the external legs} (off-mass-shell)
by the  corresponding propagators. 

We should  anticipate that the application of the NIR
 of the vertex function to the BS framework is only formal. Let us clarify
the meaning of this statement, that allows one to better understand the spirit of
extending NIR, elaborated within a perturbative framework, such as the one based on Feynman
diagrams, to a non
perturbative regime, that is compelling  for  a realistic description of  an interacting system (in a bound or
 scattering states). 
As shown in Eq. \eqref{vertex1}, the denominator contains the analytic structure
of the amplitude and the numerator is a quantity that in principle one can
obtain once a perturbation theory is considered, i.e. when a scattering process is
described perturbatively. One can tentatively assume the formal expression
\eqref{vertex1} is valid also in a non perturbative regime, but taking as arbitrary the function in the numerator and  keeping
the same analytical structure in  the denominator, i.e. the one
determined by all the independent scalar products composed by the external
4-momenta. To construct the  BSA from Eq. \eqref{vertex1} one can simply
 multiply by the constituent propagators. This step preserves
the overall 
 analytic structure shown in Eq. \eqref{vertex1} (the external momenta do not change!) but change the power of the denominator  
 (see next
 Section).

Summarizing: for positive energies of the system, one gets perturbatively 
the expression \eqref{vertex1}, and assumes that the
analytic structure is the same also for a non perturbative scenario, planning to
exploit the freedom associated to the numerator. This is taken as an
arbitrary function, once we move from the perturbative framework to the non
perturbative one.  In spirit, this strategy could resemble the familiar steps one
carries on when the harmonic oscillator eigen-problem is solved. First one determines the
asymptotic behavior of the solutions and then, one realizes that in order to 
solve the 
eigenvalue problem it is necessary the presence of the Hermite polynomials.
Similar steps are followed for actually solving  BSE once
 the NIR of the BSA is introduced.
 Noteworthy, elaborating the Nakanishi arguments \cite{nak63,nak71}, one can 
 always assume NIR for the BSA 
 if the interaction kernel in BSE  is analytical.
 
 The homogeneous BSE for two scalars has been numerically solved within NIR framework in 
 Ref. \cite{Kusa} (an early application via the uniqueness theorem of
 the Nakanishi weight function \cite{nak71}) and in Refs.
 \cite{CK2006,FSV2,Tomio2016}. It should be pointed out that also 
 the inhomogeneous BSE, suitable for
 investigating the scattering states, can be studied by using the NIR framework
 \cite{FSV1,FSV3}.


\section{Projecting  BSE onto the LF hyper-plane $x^+=0$}
 As mentioned in the previous Section, the appealing feature of  
 NIR is the explicit appearance of the analytic structure of the vertex function
 and the presence of 
   some {\em hidden} freedom, once the   weight function is taken 
  as an
   unknown quantity within a non perturbative context, i.e the BSE in
   Minkowski space. The basic question to
   be answered is: do actual solutions
   of the BSE  have  the analytic structure  exposed by NIR? In what
   follows, it will be  sketched how to reach a quantitative answer to the previous question by exploiting
    the
   LF framework (for more
   details, see Refs. \cite{CK2006,FSV2,FSV1,CK2010,dFSV1}). 
 
 If one inserts the NIR of the BSA in the homogeneous BSE, one can
 perform the needed analytic integration obtaining  an eigen-equation   for the unknown weight function (indeed one gets a
 generalized eigenvalue problem). It is the integration on the LF variable $k^-$, also
 known as  LF projection 
 (see Ref. \cite{FS} and references quoted therein  for a short 
 review of the issue), that allows one to reduce the complexity of 
  the 4D BSE.
 
 Let us illustrate the issue in the simple case of two massive 
 scalars interacting through the exchange of a massive scalar in ladder
 approximation, within the non explicitly covariant LF framework
 \cite{FSV2,FSV1} (for 
 the treatment  within the explicitly covariant LF framework see Refs.
 \cite{CK2006,CK2006b}). 
 First one  assumes a NIR for the BSA, viz
 \be
 \Phi_b(k,p)= \int_0^\infty d\gamma'\int_{-1}^1
 dz'~{g_b(\gamma',z',;\kappa^2)\over 
 \Bigl[\gamma'-{p^2\over 4}- k^2-z' k\cdot p-i\epsilon\Bigr]^3}
 \label{nir1}\ee
where  
 $g_b(\gamma',1-2\xi;\kappa^2)$ is the Nakanishi weight function, to be
 determined, $p^2=M^2$ is the 4-momentum of the interacting system and  $\kappa^2=m^2-M^2/4$ yields the strength of the binding, 
 since $m$
 is the mass of the constituents and  $M=2m-B$ with $B\equiv$ binding energy. 
 The 
 {\em  valence component} of the Fock expansion of the two-scalar interacting
 state is readily obtained  by integrating on 
 $k^-$  the BSA,  $\Phi_b(k,p)$, and reads
  \be \psi_{n=2}(\xi,k_\perp)=~{p^+\over \sqrt{2}}~\xi~(1-\xi)
\int {dk^- \over 2 \pi}
\Phi_b(k,p)=\nonu 
=~{1\over \sqrt{2}}\xi~(1-\xi)~\int_0^{\infty}d\gamma'~
{g_b(\gamma',1-2\xi;\kappa^2) \over
[\gamma'+\gamma +\kappa^2+\left(2\xi-1\right)^2 {M^2\over 4}-i\epsilon]^2}
 \ee where $\gamma\equiv |{\bf k}_\perp|^2 \in [0,\infty]$ and $\xi=(1+z)/2\in[0,1]$.
 Finally, by applying the same LF projection to both sides of the homogeneous BSE,
 written as follows 
  \be
  \Phi_b(k,p)=G_0(k,p)~ \int d^4 k'~{\cal K}_{BS}(k,k',p) ~\Phi_b(k',p)
  \label{BSE}\ee
  one gets \cite{FSV1}
  \be \int_0^{\infty}d\gamma'\frac{g_b(\gamma',z;\kappa^2)}{[\gamma'+\gamma
+z^2 m^2+(1-z^2)\kappa^2-i\epsilon]^2} =
\int_0^{\infty}d\gamma'\int_{-1}^{1}dz'~V^{LF}_b(\gamma,z;\gamma',z')
g_b(\gamma',z';\kappa^2).
 \label{nak1}\ee
with
  $ V^{LF}_b(\gamma,z;\gamma',z')$ a new kernel, fully 
  determined by the irreducible kernel 
  ${\cal K}_{BS}(k,k',p)$ present in BSE, \eqref{BSE}.
  The explicit form for the Nakanishi kernel $ V^{LF}_b$ in ladder approximation can be found in
  \cite{CK2006} and in \cite{FSV2}, while   Ref. \cite{CK2006b} presents  the cross-ladder case.
  
 In conclusion, once we know an explicit form of the 4D kernel in BSE,
  \eqref{BSE} ,then we
  can obtain a generalized eigenvalue problem for determining the Nakanishi
  weight function $g_b(\gamma',z;\kappa^2)$, as given by Eq. \eqref{nak1}.
  If solutions exist, this indicate that solutions of the BSE in Minkowski space
  can be written as a NIR. It is worth mentioning  that the same approach has been applied to 
  { excited}
  states \cite{Tomio2016}, allowing the calculation of  the valence momentum
  distributions for those states, and also to the evaluation of   scattering 
  lengths \cite{FSV3}, i.e. a first
  insight in the continuum.

\section{Spin degrees of freedom and  BSE}
  To introduce spin degrees of freedom in the BSE is a non trivial task, and one has to carefully elaborate the treatment for getting
  reliable solutions \cite{CK2010,dFSV1}. 
  An immediate consequence of the presence of further degrees of freedom is the need of decomposing the BSA 
  in a suitable sum of Dirac structures multiplied by 
 unknown scalar functions, that depend upon the external 4 momenta. Notice that
 the number of  allowed Dirac structures is
  constrained by parity, { total spin  and the proper behavior of the BSA under}
  Lorentz transformations. A first big
  difference between a two-fermion interacting system (but also for a
  fermion-boson or vector-vector cases) and a two-scalar one is  the increasing of  the number
  of scalar functions determining the BSA,  and consequently
   the number of Nakanishi weight functions.  As a matter of fact, each
  scalar function that is present in the {\em macroscopical} expansion of the 
  BSA
   can be written in terms of  a NIR, generalizing the two-scalar case. 
  
  The first case that has been investigated is a $0^+$ two-fermion
  system \cite{CK2010,dFSV1}. The corresponding BSA contains four scalar
  functions and can be written as follows
  \be
  \Phi(k,p)= S_1 ~\phi_1(k,p)+S_2 ~\phi_2(k,p)+S_3 ~\phi_3(k,p)
 +S_4 ~\phi_4(k,p)
 \label{BSE2}\ee
 where $S_i$ are the four   Dirac structures compatible with the quantum numbers of the $0^+$ state.
  For a $1^+$ state the number of $S_i$ doubles. Indeed
 there is some freedom in the actual choice of the $4\times 4$ $S_i$ matrices,
 but it is convenient \cite{CK2010} to implement an orthogonality relation
 between them. In particular, one can choose $S_i$ such that 
$ Tr \{S_i~S_j\}={\cal N}_i~\delta_{ij}$.
Hence, one gets
\be S_{1} = \gamma_5~~, \quad 
S_{2} = {\psla p\over M}  ~\gamma_5~~, \quad 
S_{3} = {k \cdot p \over M^3}  \psla p ~\gamma_5 - {1\over M} \psla k 
\gamma_5~~, 
S_{4} = {i \over M^2} \sigma^{\mu\nu}  p_{\mu} k_{\nu} ~\gamma_5
\ee
The two-fermion BSE reads 
\be 
  \Phi(k,p)=~S(p/2+k)~\int d^4k' ~F^2(k-k')~i{\cal K}(k,k')
\Gamma_1~\Phi(k',p)~\bar\Gamma_2~
S(k-p/2)
\label{BSE3}\ee
where  $S(q)$ is the Dirac propagator
 \be
S(q)=i{\psla q  +m\over  q^2-m^2+i\epsilon} \ee
and at each interaction vertex it has to be attached the following form factor
\be F(k-k')= {(\mu^2-\Lambda^2)\over  [(k-k')^2-\Lambda^2 +i\epsilon]}\ee
Finally, the Dirac structure of the interaction { vertices}
depends upon the  boson that mediates the interaction in the simple
model we are considering. In particular, in Refs. \cite{CK2010,dFSV1} three
different exchanges
have been inserted: i) a scalar boson, i.e. $ \Gamma_1=\Gamma_2= 1$; ii) 
 a pseudoscalar boson, i.e $\Gamma_1=\Gamma_2= \gamma_5$ and iii) a 
 vector boson, 
 i.e $ \Gamma_1=\Gamma_2= \gamma^\mu$. The notation $\bar \Gamma$  means
 $C \Gamma^T C^{-1}$ with $C$ the charge conjugation.
 In Eq. \eqref{BSE3}, the quantity ${\cal K}(k,k')$ is
  the  momentum-dependent part of the interaction kernel, that 
  in ladder approximation is 
given  for a massive scalar (pseudoscalar) by
   ${\cal K}_{S(PS)}=\pm{g^2/ [(k-k')^2-\mu^2+i\epsilon]}~~,$
 and  for a massless vector by
   ${\cal K}^{\mu\nu}_V={g^2~g^{\mu\nu}/ [(k-k')^2+i\epsilon]}$.
 Each unknown scalar function
  $\phi_i  $  in Eq. \eqref{BSE2} has a
   well-defined symmetry
under the exchange $1 \to 2$, as dictated by  the symmetry of both 
$\Phi(k,p)$ and the matrices $S_i$. In particular, 
$\phi_{3}(k,p)=-\phi_3(-k,p)$, while the others are symmetric. 
By using a NIR for each $\phi_i$, as in Eq. \eqref{nir1}, and applying  
a  LF projection one gets
 \be \psi_{i}(\gamma,z)=\int {dk^-\over 2\pi}~\phi_i(k,p)
=  -{i \over M} \int_0^\infty d\gamma' 
{ g_i(\gamma',z;\kappa^2) \over \left[\gamma + \gamma' + m^2 z^2 +
 (1-z^2)\kappa^2-i\epsilon\right]^2}
 \label{nir2}\ee
Finally, applying  the LF projection to the rhs of the BSE in \eqref{BSE3}, 
one can formally transform the BSE in a
 coupled-equation system, viz
\be
\psi_{i}(\gamma,z)= g^2
\sum_{j}~\int_{-1}^1 dz' \int_0^\infty d\gamma' ~
g_j(\gamma',z';\kappa^2) ~
{\cal L}_{ij}(\gamma,z,\gamma',z';p)
\label{nir3}\ee
where the matrix ${\cal L}_{ij}(\gamma,z,\gamma',z';p)$ yields the suitable
 Nakanishi kernel (its explicit expression is given in \cite{dFSV1,dFSV2}).
 From Eqs. \eqref{nir2} and \eqref{nir3}, one realizes that the coupled-equation
 system is a generalized eigenvalue problem where the Nakanishi weights
  $g_j(\gamma',z';\kappa^2)$ are the eigen-vectors, and the interaction coupling
  $g^2$ plays the role of the eigen-value, once the binding $B$ has been fixed.
      For actual calculations, 
 a suitable orthonormal basis  given by the Cartesian product of
 Laguerre polynomials in $\gamma$ and   Gegenbauer polynomials in $z$
  has been used \cite{dFSV1,dFSV2}, as in the case of the two-scalar system 
  \cite{FSV2}

\begin{table}
 \caption{Scalar couplings $g^2$ (they are  eigenvalues of Eq. \eqref{nir3}) corresponding to  a mass of the exchanged boson 
 $\mu/m=0.15$ and $\Lambda=2$ in the vertex form factor. First column: assigned binding energies. 
 Second column:  $g^2$, obtained 
 without {\em{non singular}} term in the kernel matrix. 
Third column: full calculation with {\em singular} contributions in ${\cal L}_{ij}$ exactly taken into account\cite{dFSV1}. 
Fourth column:  results obtained in Ref. \cite{CK2010} by using a smoothing function for treating 
the singular behavior of the integrals. }
 \begin{center}
 \label{tab1}
 \begin{tabular}{llll}
 \hline\noalign{\smallskip}
 B/m& $g^2_{dFSV}$(ns)&$g^2_{dFSV}$(full)\cite{dFSV1}  &$g^2_{CK}$\cite{CK2010}\\
 \noalign{\smallskip}\hline\noalign{\smallskip}
 0.01& 7.879&7.844 &7.813\\
 0.02&10.109& 10.040&10.05\\
 0.03&12.041&11.930 &11.95\\
 0.04&13.837&13.675 &13.69\\
 0.05&15.558&15.336 &15.35\\
 0.10&23.745&23.122&23.12\\
 0.20&40.738&38.324 &38.32\\
 0.30&61.449&54.195 &54.20\\
 0.40&87.303&71.060 &71.07\\
 0.50&121.342& 88.964 &86.95\\
\noalign{\smallskip}\hline
 \end{tabular}
 \end{center}
 \end{table}
 
While in the scalar  interacting system the Nakanishi kernel is regular, for the
fermionic case \cite{dFSV1,dFSV2}
the kernel  matrix ${\cal L}_{ij}(\gamma,z,\gamma',z';p)$ contains singular
contributions produced by integrating on $k^-$ the combination of  the
  numerator of the fermionic propagators in the rhs  and   the operators $S_i$
 in  $\Phi(k,p)$, on the lhs (notice that the combination is produced when we
 single out each scalar function to get Eq. \eqref{nir3}). Noteworthy,
the non explicitly covariant LF framework allows one to
straightforwardly
 determine the 
singular contributions to  ${\cal L}_{ij}$, that have the following general
form
\be
{\cal C}_j
=
\int_{-\infty}^{\infty} {dk^-\over 2 \pi}
 (k^-)^j ~{\cal S}(k^-,v,z,z',\gamma,\gamma')
 \ee
 with {  $j=1,2,3$} and 
  ${\cal S}(k^-,v,z,z',\gamma,\gamma')$  explicitly calculable
  \cite{dFSV1,dFSV2}.
  For some values of the variables $\{v,z,z'\}$, one can have  the worst case 
  $
  {\cal S}(k^-,v,z,z',\gamma,\gamma')\sim {1/ {  [k^-]^2}}$ 
  for $k^- \to \infty$
     Then, one cannot close the arc at  $\infty$ 
  for carrying out
  the needed analytic integration, but has to deal with singular behavior. Fortunately this kind of LF singularities can be carefully
  treated \cite{dFSV1,dFSV2} by exploiting the studies of the field theory in 
  the
    Infinite Momentum
frame, performed in the seventies by T. M. Yan and collaborators 
(see in particular Ref. \cite{Yan}).
The relevant  singular integral has the following  expression 
  \be{\cal I}(\beta,y)=\int_{-\infty}^\infty 
{ dx \over \Bigl[\beta x -y\mp i \epsilon\Bigr]^2}= \pm 
{2\pi i~\delta (\beta)\over \Bigl[-y\mp i \epsilon\Bigr]}\ee
Indeed, to fully account the singular behavior encountered in our analysis it is also needed to consider 
$1/2 \partial/\partial \beta {\cal
I}(\beta)$ that means to deal with $\partial/\partial \beta~\delta(\beta)$. From the numerical point of view, this does not represent  an
issue, given the orthonormal basis adopted for expanding the Nakanishi weight functions.

Another positive fact is given by recognizing that in ladder approximation
  the severity of the singularities, i.e.   the power $j$,  depends
   only  
upon  the constituent propagators and the
structure of the BSA.
 Differently,  in the explicit covariant LF framework, 
the trouble produced by the singular behavior of the relevant integrals needed for solving BSE with NIR   was pragmatically 
fixed by introducing 
a suitable smoothing function \cite{CK2010}.
 
An example of the numerical results we have obtained, and the quality of the comparisons we have achieved is given in Table \ref{tab1},
where for an exchanged scalar  with mass $\mu/m=0.15$, the values of the coupling constant $g^2$  corresponding to different values of the
binding energy of the two-fermion system are given.  The outcomes are obtained by solving the generalized eigenvalue problem in Eq.
\eqref{nir3}, after discretazing the coupled-equation system by expanding the Nakanishi weight functions $g_i(\gamma,z;\kappa^2)$ on the orthonormal
basis above mentioned. The effect generated by the LF singularities can be appreciated  by comparing the second column (without singular
contributions) and the third one \cite{dFSV1} (with  singular contributions). In order to have a complete overview of the issue,  the fourth 
column  shows the results obtained in Ref. \cite{CK2010} by introducing a smoothing function for taking numerically under control the
 plague of the singularities. A more exhaustive comparison between the results obtained within the non covariant LF framework and the ones
 in the explicitly covariant LF approach is presented in Ref. \cite{dFSV1}.  The very nice agreement between results obtained in Minkowski space 
 is made complete by 
 the favorable comparison, also  shown in \cite{dFSV1}, with the results 
{ from   Euclidean space calculations} \cite{Dork}.
 
The appealing 
{ to get}
solutions 
directly in Minkowski space is given by the possibility to evaluate LF 
 distributions, as shown in
 \cite{dFSV1}.  There, the dynamical effect of the ladder exchange is illustrated by looking at 
 the tail of  the transverse-momentum behavior of the amplitude $\psi_i(\gamma,z)$ that nicely results 
 in agreement
 with the finding in Ref. \cite{Ji2003}, where the asymptotic behavior of 
  $\psi_i$ for the pion 
 (cf  Eq. \eqref{nir2})  is  determined by exploiting a very general counting rule.

\section{Conclusions \& Perspectives}

The technique for solving
the fermionic  BSE in Minkowski space by using the Nakanishi integral representation of the Bethe-Salpeter amplitude
can be now safely applied and extended, since a crucial point related to the treatment of light-front singularities has been singled out and
fixed.  The 
   rule for dealing with the expected singularities, that in ladder
   approximation
  basically depend upon the structure of the BSA and not upon the complexity of the
kernel, has been established \cite{dFSV1,dFSV2} and a short discussion has been illustrated  in the present  contribution. 
To reach a satisfactory treatment of the above mentioned singularities,  that allows one to open the possibility of 
investigating a wide set of    interacting systems with spin degrees of
 freedom, the LF framework plays a basic role due to the
  well-known advantages in performing  
analytical
integrations. In ladder approximation, after obtaining a manageable form  of the eigenvalue problem for the $0^+$ system of two fermions,
we have solved  the coupled-equation system of four integral equations for the Nakanishi weight functions by using a suitable orthonormal basis. 
Our numerical investigations confirm both the robustness of the Nakanishi Integral Representation
 for the BSA, valid for any analytical BS kernel,
and strongly encourages to extended the technique to other interesting cases: 
boson-fermion  and
vector-vector systems.

Calculations are in progress for  the LF momentum distributions 
of the two-fermion system in the valence component, elucidating some formal
subtleties.
 
{\it Acknowledgments.}  W. de Paula and T. Frederico thank the support of the
Brazilian Institution 
CNPq, and
 G. Salm\`e acknowledges the support by CAPES.

\end{document}